\documentclass[useAMS]{mn2e}
\usepackage{graphicx}

\title[Variable dust formation by HD 36402]{\bf Variable dust formation by the  
colliding-wind Wolf-Rayet system HD~36402 in the Large Magellanic Cloud}
\author[P. M. Williams et al.]
       {P. M. Williams$^{1}$\thanks{E-mail: pmw@roe.ac.uk},
        Y.-H. Chu$^{2}$, R. A. Gruendl$^{2}$ and M. A. Guerrero$^3$\\
   $^{1}$Institute for Astronomy, University of Edinburgh, Royal Observatory, Edinburgh EH9 3HJ\\
   $^{2}$Department of Astronomy, University of Illinois at Urbana-Champaign, 
         1002 West Green St, Urbana, IL 61801-3080, USA\\
   $^{3}$Instituto de Astrofisica de Andalucia, Glorieta de la Astronomia, s/n, 18008 Granada, Spain\\}
\date{Accepted 2013 February 6.
      Received 2013 February 6;
      in original form 2012 December 19 in text/HD36402/submit/HD36402rv.tex}

\pagerange{\pageref{firstpage}--\pageref{lastpage}}
\pubyear{2013}

\begin{document}

\maketitle

\label{firstpage}

\begin{abstract}
Infrared photometry of the probable triple WC4(+O?)+O8I: Wolf-Rayet system HD~36402 
(= BAT99--38) in the Large Magellanic Cloud (LMC) shows emission characteristic of heated dust. 
The dust emission is variable on a time-scale of years, with a period near 4.7~yr, possibly 
associated with orbital motion of the O8 supergiant and the inner $P\simeq3.03$-d WC4+O binary. 
The phase of maximum dust emission is close to that of the X-ray minimum, consistent with 
both processes being tied to colliding wind effects in an elliptical binary orbit.
It is evident that Wolf-Rayet dust formation occurs also in metal-poor environments.
\end{abstract}


\section{Introduction}
Population I Wolf--Rayet (WR) stars lose mass at prodigious rates ($> 10^{-5}$M$_{\odot}$yr$^{-1}$) 
in dense, fast stellar winds giving the stars their characteristic emission-line spectra: they 
represent the last stable stage in the evolution of massive stars. Their high luminosities and 
conspicuous spectra facilitate identification in external galaxies, while the stellar winds have 
a profound influence on the local interstellar medium, enriching it with nuclear-processed 
material and frequently forming bubbles. 
When WR stars are members of binary systems with other luminous stars possessing stellar winds, 
the collision of the two stellar winds can result in shock-heating of stellar wind material and 
particle acceleration, observable in X-ray and non-thermal radio emission. Some colliding-wind 
binaries (CWBs) incorporating WC type WR stars also form carbon dust in their winds, observable by 
its conspicuous infrared (IR) emission. These phenomena often vary periodically, especially if 
the stars move in an elliptic orbit and their separation varies so that the wind-collision 
region (WCR) moves periodically between denser or rarer regions of the stellar winds. 
Also, depending on the inclination of the system, our line of sight to the WCR passes through 
varying amounts of stellar wind as the orbit progresses, causing variations in the free-free and 
photoionization extinction to the observed radio and X-ray emission. 
The prototype is WR\,140 (= HD~193793), whose IR, radio and X-ray variations were linked to 
the motion of its WC7 and O5 components in their 7.94-yr orbit by Williams et al. (1990). 
This system has continued to attract observations in all wavelength domains -- for a recent 
summary, see Williams (2011a) -- as a laboratory of particle acceleration 
(e.g. Pittard \& Dougherty 2006) and led to the identification of other variable dust-making 
WR CWBs in the Galaxy. The observation of rotating dust `pinwheels' around some of the WR 
systems which show persistent dust emission (e.g. WR\,104, by Tuthill et al. 2008) associates 
these with CWBs too. The persistent dust makers are nearly all WC9 stars and their locations show 
a strong preference for the metal-rich region of our Galaxy, generally towards the Galactic Centre. 
On the other hand, the variable and episodic WR dust makers have spectral types ranging from WC5 
to WC8 and a less concentrated galactic distribution -- but all of them are located within the 
Solar circle. This suggests a strong influence of environmental metallicity on the phenomenon of 
dust formation by WR systems.

In the lower metallicity environments of the Magellanic Clouds, the fractions of WR/O stars 
and distributions of WR spectral subtypes are very different from those in our Galaxy. In 
particular, of the 24 WC/WO stars in the LMC, 23 are of type WC4 and one a WO3 (Moffat 2008). 
The question arises: do any of these make carbon dust? This can be tackled by a survey of 
their spectral energy distributions (SEDs) or a search for variability in the IR 
characteristic of the variable or episodic dust-makers like WR\,140. 
To begin, we examined the $Ks$-band photometry of the LMC WC4 stars measured in the Deep 
Near Infrared Survey of the Southern Sky (DENIS; Epchtein et al. 1999, Cioni et al. 2002), 
Two Micron All Sky Survey (2MASS, Skrutskie et al. 2006) including the 6X w/LMC/SMC Point 
Source Working Database, and InfraRed Survey Facility (IRSF) Magellanic Cloud 
(Kato et al. 2007) surveys for evidence of variability. Of the 21 LMC WC4 stars with at least 
three observations, the best candidate for variability, after discounting stars with uncertain 
data or lying in very crowded regions where close companions could be included in the apertures 
of some but not other surveys, was HD~36402 (= BAT99--38, Breysacher, Azzopardi \& Testor 1999).  
It would certainly be worth repeating 
the search for variability when the VISTA Magellanic Clouds Survey (Cioni et al. 2011), with its 
higher resolution and homogeneity, is complete. The case for HD~36402 being a WR dust maker 
was strengthened by its mid-IR (3.6--8 $\umu$m) photometry in the {\em SPITZER} SAGE LMC survey 
(Bonanos et al. 2009), where it was identified as the reddest WR star. A preliminary account 
of the present study was presented at the 2010 LIAC 
(Williams 2011b, Paper~I), where it was shown that the 2--8 $\umu$m SED was characteristic of 
of WR dust, but that at wavelengths beyond 10~$\umu$m, the emission 
from the WR star could not easily be resolved from that by two nearby luminous young stellar 
objects (YSOs) discovered by Chu et al. (2005) and that the 70- and 160-$\umu$m fluxes tentatively 
ascribed to a ring nebula associated with HD 36402 by Bonanos et al. came entirely from the YSOs.

Independently, Boyer et al. (2010) came to a similar conclusion in their study of the far-IR 
emission from three dust-making evolved stars in the LMC with {\em Herschel} instrumentation. 
They presented the SEDs of HD~36402 and of a strong far-IR source, HD~36402~IR1 immediately 
to the west of the WR star, having $T_{\rmn d} \simeq$ 64 and 8~K. 
They considered the possibility that IR1 could be heated by the stellar flux from HD~36402 
but concluded that it was more likely to be the YSOs embedded in a molecular cloud.

\begin{figure}                                                          
\centering
\includegraphics[angle=270,width=8cm]{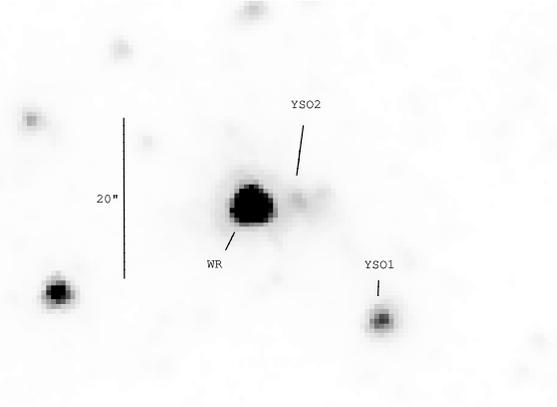}  
\caption{Field of HD 36402 (marked `WR') from the 3.6-$\umu$m IRAC image, with 
North at the top, East to the left. The YSOs discussed by Chu et al. are marked 
`YSO1' and `YSO2'. The separation of YSO-2 from the WR star is 7 arcsec and it 
is 4.3 mag. fainter than the WR star at 3.6~$\umu$m and 1.7 mag. fainter at 8.0~$\umu$m.}
\label{Fchart}
\end{figure}

The field is certainly crowded at longer wavelengths, making interpretation of long-wavelength 
observations with smaller instruments (e.g. {\em IRAS} or {\em MSX}) with their larger beam 
sizes difficult, but there is no confusion in the IRAC bands -- even at 8 $\umu$m, HD~36402  
is 1.7 mag.\, brighter than the closer YSO (YSO-2, 7~arcsec to the W of the WR star). 
At this wavelength, the IRAC beam (Fazio et al. 2004) is smaller than 2 arcsec full width half 
maximum (FWHM), so we consider YSO-2 to be well separated from HD 36402 in all IRAC bands 
as the magnitude differences are greater at shorter wavelengths. 
The field near HD~36402 at 3.6~$\umu$m is shown in Fig.\,\ref{Fchart}.
  
The spectrum and orbit of HD~36402 were studied by Moffat, Niemela \& Marraco (1990), 
who derived an orbit from the WR emission lines having a period of 3.03269~d and radial 
velocity (RV) amplitude $K = 275\pm17$~km~s$^{-1}$. The absorption-line RV amplitude 
was not significantly different from zero ($K = 8\pm9$~km~s$^{-1}$), suggesting that 
the O8 supergaint did not participate in the 3-d orbit and that HD~36402 was a triple 
system like $\theta$~Mus (= WR\,48) or R~130 in the LMC. 
From five high-resolution {\em IUE} spectra, Konigsberger, Moffat \& Auer (2003) 
measured RVs of photospheric absorption lines which fitted a sinusoidal variation 
in anti-phase with the emission-line spectrum, suggesting formation in the close 
companion to the WC star, but more probably produced by the superposition of the 
stationary absorption lines from the O8 supergiant and rotationally broadened 
absorption lines produced in the close companion. 
In their study of HD~193793 (WR\,140), Fitzpatrick, Savage \& Sitko (1982) compared 
its {\em IUE} spectrum with that of HD~36402. The $\lambda$1550 C\,{\sc iv} profile in 
HD~36402 showed narrow interstellar lines from both the Milky Way and the LMC and two 
sets of circumstellar absorption lines. The broader set was shifted $-3100$~km~s$^{-1}$ 
relative to the LMC interstellar lines and the narrower pair by $-1300$~km~s$^{-1}$. 
The same features, at the same velocities and strengths, were seen in another {\em IUE} 
spectrum taken three months earlier, ruling out their formation in a transient event. 
Niedzielski \& Sk\'orzy\'nski (2002) used the {\em IUE} high-resolution spectra to 
determine a terminal velocity of $3072\pm160$ km~s$^{-1}$, comparable to those of 
other WC4 stars in the LMC and our Galaxy. 
It seems reasonable to attribute the $-1300$ km~s$^{-1}$ absorption to the O8I 
component; it is in line with the range of terminal velocities (955--2186 km~s$^{-1}$) 
measured for O8 supergaints by Prinja, Barlow \& Howarth (1990).

In this paper, we present new observations of HD~36402 from images taken for the N51D study 
by Chu et al. and post-cryo observations extracted from the {\em Spitzer} Heritage Archive. 
Recently, additional mid-IR observations have become available from the {\em AKARI} 
LMC (Kato et al. 2012) and {\em WISE} (Wright et al. 2010) surveys. We use these data 
to re-examine the IR variability of HD 36402 and derive a period, and consider its status 
as a dust-making WR CWB.

\section{Observations}

\subsection{IR Photometry of HD 36402 from Surveys}
\label{SSurveys}

\begin{table} 
\caption{Near-infrared photometry of HD 36402}
\begin{center} 
\begin{tabular}{lcccc}
\hline 
Survey or source   & Date    & $J$   & $Ks$  & $\sigma Ks$\\
\hline
DENIS strip 4963   & 1996.92 & 11.63 & 10.55 & 0.02 \\
DENIS shifted      &         & 11.73 & 10.69 & 0.02 \\
2MASS all-sky      & 2000.08 & 11.76 & 10.97 & 0.02 \\
2MASS 6X PSWDB     & 2000.98 & 11.73 & 10.82 & 0.02 \\
2MASS 6X PSWDB     & 2001.09 & 11.69 & 10.73 & 0.02 \\
IRSF LMC0525-6720A & 2002.88 & 11.80 & 11.22 & 0.01 \\
N51D study (ISPI)  & 2006.85 & 11.73 & 10.75 & 0.05 \\
\hline
\label{TnIR}
\end{tabular} 
\end{center} 
\end{table}

\begin{table*}
\caption{Mid-IR photometry of HD 36402 with {\em Spitzer} IRAC from 
the N51D study, the SAGE IRAC Epoch 1 and Epoch 2 Catalog and archive; 
the {\em AKARI} LMC Survey Catalogue and the {\em WISE} All-Sky and 
post-cryo surveys. The {\em AKARI} $S11$ and {\em WISE} $W3$mpro 
11-$\umu$m magnitudes are given in the same column.}
\begin{center}
\begin{tabular}{lcccccccccl}
\hline
Date    &    $N3$       &    $W1$mpro   &  [3.6]        &     [4.5]     &  $W2$mpro     &     [5.8]     &     $S7$     &    [8.0]       &  $S11$/$W3$ & Source    \\
        & (3.2 $\umu$m) & (3.35 $\umu$m)& (3.55 $\umu$m)& (4.44 $\umu$m)& (4.60 $\umu$m)& (5.58 $\umu$m) & (7.0 $\umu$m)& (7.76 $\umu$m)  & (11 $\umu$m) &       \\
\hline
2004.96 &               &               & 9.71$\pm$0.05 & 8.99$\pm$0.04 &               & 8.54$\pm$0.04 &               & 7.91$\pm$0.03 &              & N51D \\ 
2005.55	&               &               & 9.30$\pm$0.05	& 8.70$\pm$0.04	&               & 8.21$\pm$0.03	&               & 7.81$\pm$0.02 &              & SAGE1 \\ 
2005.82	&               &               & 9.18$\pm$0.05	& 8.50$\pm$0.04	&               & 8.11$\pm$0.03	&               & 7.67$\pm$0.03 &              & SAGE2 \\ 
2007.42 & 9.68$\pm$0.04 &               &               &               &               &               & 7.70$\pm$0.02 &               & 7.11$\pm$0.02& LMCC\\
2009.77 &               &               &               & 8.96$\pm$0.05 &               &               &               &               &              & IRAC \\ 
2009.91 &               &               & 9.51$\pm$0.05 &               &               &               &               &               &              & IRAC \\ 
2010.41 &               & 9.44$\pm$0.02 &               &               & 8.57$\pm$0.02 &               &               &               & 7.15$\pm$0.02& {\em WISE}\\
2010.91 &               & 9.20$\pm$0.03 &               &               & 8.31$\pm$0.02 &               &               &               &              & {\em WISE}\\ 

\hline
\label{TmidIR}
\end{tabular}
\end{center}
\end{table*}

\begin{figure}    
\centering
\includegraphics[angle=270,width=8.5cm]{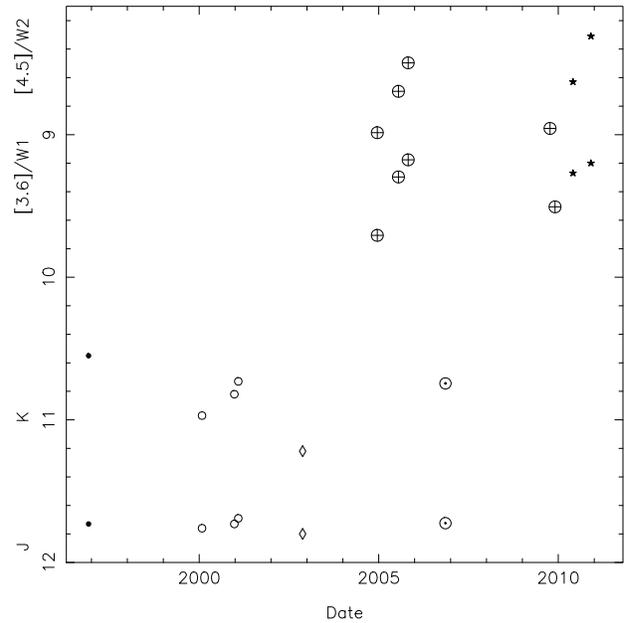}
\caption{Synoptic photometry of HD 36402 from different sources: $J$ and $Ks$ magnitudes 
from the DENIS (marked $\bullet$), 2MASS ($\circ$) and IRSF ($\diamondsuit$) surveys, 
and the ISPI frames observed for the N51D study ($\odot$); 
[3.6] and [4.5] from IRAC ($\oplus$), and $W1$ and $W2$ from {\em WISE} ($\star$).}
\label{Fsynoptic}
\end{figure}

Near-IR photometry and dates of observation of HD 36402 extracted from the DENIS, 
2MASS and IRSF Magellanic Cloud surveys are collected in Table \ref{TnIR} 
and plotted in Fig.\,\ref{Fsynoptic}. For an object having the $J-Ks$ colour 
of HD~36402, the difference between the IRSF and 2MASS $Ks$ magnitudes is 
not expected to exceed 0.02 mag. (Kato et al. 2007, table 10). 
The differences between DENIS and 2MASS magnitude scales are, however, greater. 
Delmotte et al. (2001) measured systematic shifts between the $J$ and $Ks$ 
scales of the two catalogues, finding means $\delta$ = --0.10 ($J$) and --0.14 
($Ks$) in the sense DENIS minus 2MASS, but with significant variation from 
DENIS strip to strip. 
These differences are applied to the data in Table \ref{TnIR} 
and Fig.\,\ref{Fsynoptic} to align the DENIS and 2MASS magnitudes.

Mid-IR photometry of HD 36402 is available from the {\em Spitzer}, {\em AKARI} 
and {\em WISE} missions. 
Magnitudes of HD~36402 in the {\em Spitzer} SAGE survey of the Magellanic Clouds 
(Meixner et al. 2006) extracted from the NASA/IPAC Infrared Science Archive are 
given in Table \ref{TmidIR} and the [3.6] and [4.5] magnitudes are plotted against 
date of observation in Fig.\,\ref{Fsynoptic}. 

HD 36402 was observed in the {\em AKARI} LMC Survey (Kato et al. 2012) in the 
$N3$ ($\lambda$ 3.2-$\umu$m), $S7$ ($\lambda$ 7.0-$\umu$m) and $S11$ 
($\lambda$ 11-$\umu$m) bands as AKARI-LMCC J052603.96-672956.7 on 2007 June 2--3. 
The data are included in Table \ref{TmidIR}.  

Owing to its high ecliptic latitude (87$\degr$), HD~36402 was observed very 
frequently by {\em WISE} in both the All-sky and Post-Cryo surveys. 
It was flagged as having a high probability of variability in the All-Sky survey, 
so we examined this using the 173 `A'-quality $W1$ (profile-fitted, 3.4-$\umu$m) 
magnitudes extracted from the {\em WISE} All-Sky Single Exposure (L1b) Source Table.
The data covered 21 days (MJD 55335--356 in 2010 May-June) and we began by 
searching for periodicity using the {\sc clean} algorithm 
(Roberts, Leh\'ar \& Dreher 1987) as implemented in the 
Starlink\footnote{The Starlink software suite is currently maintained and 
distributed by the Joint Astronomy Centre, Hilo, Hawaii.} {\sc period} package 
(Dhillon, Privett \& Duffey 2001). We found a peak at 5.99$\pm$0.04~d but the 
false alarm probability (FAP) calculated from 250 random permutations of the 
data was 0.056 ($\sigma$ = 0.015), so the period is only marginally significant. 
We repeated the period search on the 202 $W1$ magnitudes observed during 
MJD 55522--539 (2010 November -- 2011 January from the {\em WISE} Preliminary 
Post-Cryo Single Exposure (L1b) Source Table, but found no significant period, 
so we discount that suggested by the All-Sky data. 
We also examined the data for a trend and found the $W1$ magnitudes to be 
brightening at 0.0023$\pm$0.0007 mag~d$^{-1}$ during the 21 days of 
observations in the All-Sky survey. This is consistent with the brightening 
in $W1$ and $W2$ magnitudes between the All-Sky and post-cryo 
surveys (Table \ref{TmidIR} and Fig.\,\ref{Fsynoptic}), where the dates used 
are the average dates derived from the Source Tables. These show $W1$
brightening at an average rate of 0.0013$\pm$0.0002 mag~d$^{-1}$, comparable 
to the rates of the IRAC 3.6- and 4.5-$\umu$m fluxes. 

At 11-$\umu$m the beams of the {\em AKARI} $S11$ and {\em WISE} $W3$ magnitudes 
are 5.5 and 6.5 arcsec FWHM respectively and we hestitated to use these data for 
HD 36402 in case of possible contamination from YSO2, whose spectrum is rising 
in these relatively wide (4.6 and 8 $\umu$m respectively) filter pass-bands while 
that of HD 36402 is expected to be falling. 
But, as seen from the consistency of these fluxes with the SEDs fitted to the 
shorter wavelength data in Section \ref{SSED} below, most if not all of the flux 
measured in the $S11$ and $W3$ bands can be attributed to the WR star.

\subsection{New photometry of HD 36402}

In their observation of the N51D region, Chu et al. (2005) drew attention to the 
bright image of HD~36402 close to YSOs they reported. As the IRAC observations 
for this study were taken on 2004 December 12, before the SAGE observations, we 
measured the flux from HD 36402 in the IRAC images to extend knowledge of its 
variability. The results are included in Table \ref{TmidIR}.  

The region was observed in the near-IR with the Cerro Tololo Inter-American 
Observatory (CTIO) Blanco 4.0-m Infrared Side Port Imager (ISPI) on 2006 November 7. 
Three fields including HD 36402 were observed and the photometric calibration was 
tied to 2MASS observations of stars in the fields. The mean $J$ and $Ks$ magntudes 
are included in Table \ref{TnIR}. 

Thirdly, the {\em Spitzer} Heritage Archive was searched for images of fields 
including HD 36402. Twelve observations in [4.5] during 2009 October 3--15 
and eleven in [3.6] between 2009 November 16 and December 14 were retrieved and 
magnitudes of HD 36402 measured. The WR star was generally near the edge of the 
field, very close to the edge in the case of the [3.6] observations, making sky
estimation difficult. Both bands show brightening during the observation 
sequences but not at statistically significant rates. The average magnitude and 
date for each of the two runs is given in Table \ref{TmidIR} and plotted in 
Fig.\,\ref{Fsynoptic}.

\section{Results}
\subsection{Light curves and periodicity}

\begin{figure}                                         
\centering
\includegraphics[width=8.5cm]{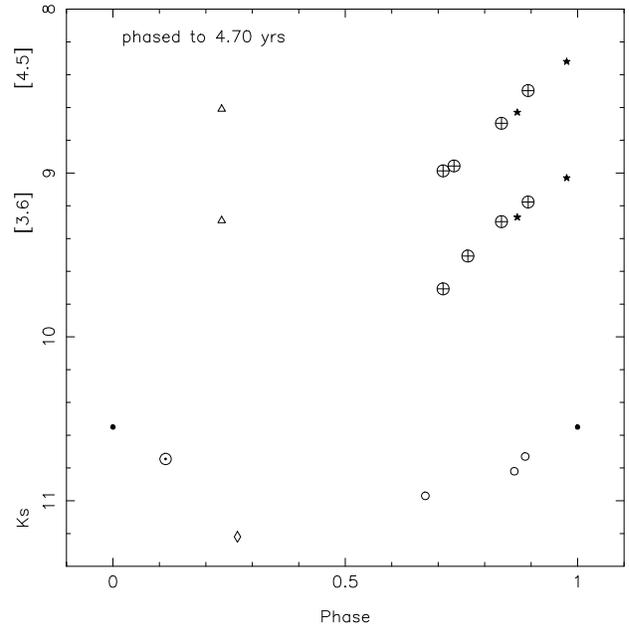}
\caption{Photometry of HD 36402 phased to a period of 4.7~yr with zero phase 
set to the epoch of the DENIS observation. For plotting against phase, 
[3.6] and [4.5] magnitudes have been interpolated from the {\em WISE} $W1$ and 
$W2$ and the {\em AKARI} LMCC $N3$ and $S7$ magnitudes as described in the text. 
Points from the {\em AKARI} data are marked $\triangle$, while the sources of the 
other data are as in Fig.\,\ref{Fsynoptic}. }
\label{Fphased}
\end{figure} 

It is evident from the IR observations plotted in Fig\,\ref{Fsynoptic} that the 
amplitude of variation increases with wavelength. This is characteristic of 
varying circumstellar dust emission by WR stars, which peaks in the 
3--4 $\umu$m region, and reflects  the increasing contribution of the dust 
emission with wavelength, which can be seen in the SED (Fig.\ref{FSEDs}, Paper~1, 
and Boyer et al.). 

Variation in the dust emission can arise from variation in the stellar flux 
heating the dust or in the amount of circumstellar dust. We observe very 
little variation in $J$ ($\sigma$ = 0.04 mag., Fig.\ref{Fsynoptic}) because the 
dust contributes little to the flux at this wavelength. The near constancy also 
suggests that the stellar radiation dominating the $J$-band flux is not 
significantly variable. There is some dispersion in the visible photometry of 
HD 36402 -- e.g. $V$ = 11.44 (Feitzinger \& Isserstedt 1983), 11.618$\pm$0.098 
(Zaritsky et al. 2004) -- 
but a concerted search for variability by Seggewiss, Moffat \& Lamontagne (1991) 
showed no variation greater than the photometric uncertainties. We therefore 
deduce that the variation in the IR comes mostly or wholly from changes in the 
amount of dust in the wind rather than the stellar radiation heating it.
The dust forms in the shock-compressed stellar wind and inherits its velocity, 
moving away from the stars as seen in the pinwheels made by systems like WR\,104, 
whose winds are relatively slow, or the expanding splash made by WR\,140 
(Williams et al. 2009), whose wind is 2--3 times faster. As the dust moves 
further from the stars, the stellar radiation heating it is progressively 
diluted and the grains cool, so that their emission fades. Therefore, if a system 
is observed to show constant IR emission, the dissipating dust must be constantly 
being replenished by the condensation of new dust grains -- the observed IR SED is 
a consequence of the balance of these two processes. The growth of IR emission by 
HD 36402 in 2004--05 and 2009--10 indicates that the rate of dust formation must 
have been increasing for over a year in each case. Between these events, the rate 
of condensation must have fallen below the rate needed to replenish the dissipation 
to account for the fade in emission between the two rises.

The similarity of the rates of brightening in 2004--05 and 2009--10 suggests that 
the variation is periodic. To facilitate comparison of the {\em WISE} and IRAC 
data, we interpolated [3.6] and [4.5] magnitudes from the 
{\em WISE} $W1$ (3.4-$\umu$m) and $W2$ (4.6-$\umu$m) magnitudes by fitting black 
bodies to the $W1$ and $W2$ fluxes, calculating fluxes at 3.6 and 4.5 $\umu$m, 
and converting them to magnitudes. Comparison of these with the IRAC observations 
shows an interval of 4.7$\pm$0.2~yr between the two rises. The uncertainty was 
estimated from inspection of trial fits, allowing for the spreads in epochs of the 
{\em WISE} observations, and is probably conservative. 
In Fig.\,\ref{Fphased}, we show the $Ks$, [3.6] and [4.5] magnitudes phased to the  
period of 4.7~yr, with zero phase arbitrarily set to the epoch of the DENIS observation.
The 2000--01 2MASS $Ks$ photometry fitted to this ephemeris shows a brightening towards 
a possible maximum near the time of the DENIS observation parallelling that seen in the 
mid-IR, supporting the interpretation of the 4.7-yr shift as a period in the variations 
as the 2MASS data come from an earlier cycle than the mid-IR data.

The IRAC and {\em WISE} data can also be fit by the first harmonic (2.35 yr) of this 
period, but the [3.6] and [4.5] magnitudes determined from interpolating the {\em AKARI} 
$N3$ and $S7$ observations by fitting a black body to them do not fit, so we discount 
this possibility. Higher harmomics are ruled out by introduction of scatter in the 
phased near-IR data or sequence of the IRAC data, so we conclude that the variations 
are best fit with a period near 4.7 yr.

For the present discussion, we adopt a preliminary ephemeris for the near-IR 
maximum of 
\[
T (Y) = 1996.9 + 4.7 N
\]
\noindent Further near-IR photometry is needed is define the light curves, including 
fading, the minimum and especially the rise to the next maximum expected in 2015.7 to 
confirm the period.

The phase of the {\em AKARI}-derived [3.6] and [4.5] magnitudes, also plotted in 
Fig\,\ref{Fphased}, places them on the falling branch of the light curve. Also, the 
phase of the 2006 ISPI $Ks$ magnitude fits the fade to the faint $Ks$ observed by IRSF. 
Whether the IRSF point represents the true minimum is not certain: the IRSF $J$--$Ks$ 
colour (0.58) is close to the mean $J$--$Ks$ (0.55$\pm$0.07) observed for other LMC WC4 
systems, so there may be no dust contribution at that phase. This would imply that 
condensation of new dust grains would have ceased some time earlier to have allowed the 
existing dust to have dissipated and its emission to have faded sufficiently. 

The light curves are not yet well defined but it is evident that the rise to maximum 
takes at least 1.5 yr. In this respect, HD 36402 more closely resembles the Galactic 
WC7+O9 dust maker, WR\,137 (HD 192641), whose IR flux takes about 3~yr to rise to 
maximum (Williams et al. 2001), than the prototype, WR\,140, which shows a much faster 
rise (less than two months in $K$, Williams et al. 2009). Better definition of the IR 
maximum from further observations will show if it occurs later at longer wavelengths 
owing to the effects of grain growth after condensation (cf. WR\,140).

\subsection{Spectral energy distributions (SEDs)}
\label{SSED}

\begin{figure}                                                              
\centering
\includegraphics[width=8.5cm]{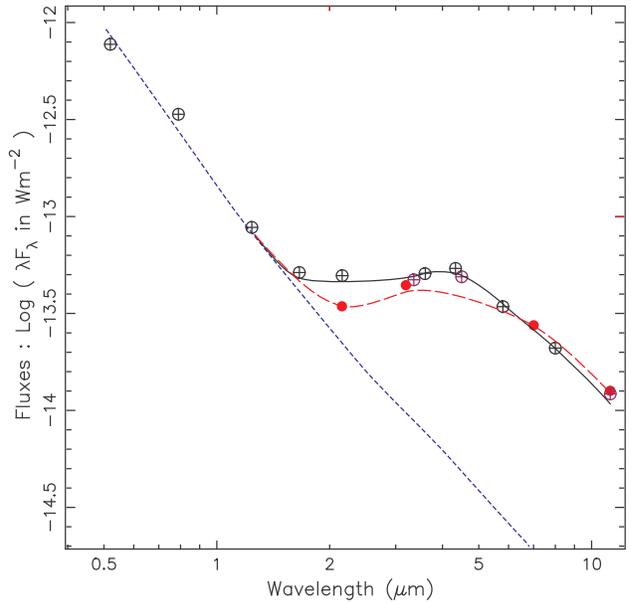}
\caption{SEDs of HD 36402 near photometric phases 0.9 and 0.2. The dotted line 
(blue in the on-line version) represents the stellar wind spectrum fitted to $v$ 
and $I$ and determined from that of the dust-free system HD 35517. 
The $\phi \simeq$ 0.9 SED based on the 2001.09 2MASS, 2005.82 IRAC and 2010.41 
{\em WISE} data is marked $\oplus$. 
A simple model adding a best-fit black body to the wind emission is shown with a solid line. 
The $\phi \simeq$ 0.2 SED, based on {\em AKARI} and interpolated $J$ and $Ks$ photometry is 
marked $\bullet$, and the corresponding model is drawn with a dashed line (red on-line).}
\label{FSEDs}
\end{figure}

To study the SED of HD 36402, we need observations over a wavelength interval 
wide enough to cover the flux maximum, essentially contemporaneous near- and 
mid-IR photometry. We do not have such data, but note that the interval between 
the 2005.82 IRAC and 2001.09 2MASS observations was 4.73 yr, and that between the 
2010.41 and the same 2MASS observations was 9.32 yr, close to our proposed period
or double it. Assuming that the SED repeated with this period, we formed a 
SED combining these data to represent the system near phase 0.9 and present it 
in Fig.\,\ref{FSEDs}. For a 4.7-yr period, the two sets of mid-IR data straddle 
the near-IR data; when the period is revised with more near-IR data, it should 
be possible to select more closely matched mid- and near-IR data. 
Owing to the breadth of the $W3$ filter, the colour correction for calibration of 
monochromatic flux from the in-band magnitude can be significant and we adopted 
that for a 800-K black body from Wright et al. (2010, table 1); those for the other 
filters are less than 2~per cent. 

For the shorter wavelengths, we used the narrow-band $v$ observed by Smith (1968), 
which avoids the strong WC emission lines falling in the $V$ band, and $I$ from 
the DENIS survey.  Smith also derived a colour excess of $E_{b-v} = 0.03$, 
which seems low considering the location of HD~36402, but is consistent with the 
reddening, $E(B-V) = 0.00\pm0.03$, derived by Morris et al. (1993) from nulling 
the 2200-\AA\ interstellar absorption feature. We de-reddened the photometry using 
Smith's $E_{b-v}$, equivalent to $A_V = 0.11$, and a standard reddening law, but 
the effects are too small to influence the results significantly.
For the stellar plus wind continuum, we noted that the SED of the LMC WC4+O5--6 
system HD 35517 (BAT99-28) defined by the 2MASS, IRAC and {\em WISE} data has 
a power-law form, with no evidence of dust emision. We shifted this to fit the 
short-wavelength photometry to provide the wind SED shown in Fig.\,\ref{FSEDs}.
The difference between the phase 0.9 and wind SEDs can be fit by a black body 
-- not as a physical model, but as a simple characterisation of the dust emission. 
A least-squares fit gave $T_{\rmn bb} = 980\pm20$~K, with r.m.s. residuals 
of 0.07 mag. The error would be greater if the near- and mid-IR photometry did 
not match, i.e. if the system does not repeat with a period of 4.7 yr. 

As noted above, we were hesitant to include the 11-$\umu$m data in case of 
contamination from YSO2. The modelling was therefore made with and without the 
$W3$ magnitude. The difference in fitted $T_{\rmn bb}$ was less than 20~K, showing 
that the $W3$ datum is consistent with the SED determined at shorter wavelengths 
and that most, if not all, of the measured $W3$ flux comes from the WR star.

To examine the maximum dust emission observed, we formed a composite SED from the 
DENIS (shifted as above) and 2010.91 {\em WISE} observations, which were taken 
2.98 cycles apart. This lies a little above the SED in Fig.\,\ref{FSEDs}, but 
is omitted for clarity. From it we measure the dust luminosity to be 
$8 \times 10^{-14}$ W~m$^{-2}$. We can compare this with the total luminosity 
of the three stars heating the dust. For the O8I star we adopt  
$\log L/L_{\odot} = 5.68$ following Martins, Schaerer \& Hillier (2005). 
We note the unseen companion to the WC4 star in the 3-d orbit would have a 
luminosity comparable to that of the O8 supergiant if it were an O4 
main-sequence star, but would be $\sim 0.8$ mag fainter in the visible -- thereby 
accounting for its not being observed in the spectra taken for RV study by 
Moffat et al. 
For the WC4 star, we assume the mean luminosity from the study of LMC WC4 stars 
by Crowther et al. (2002): $\log L/L_{\odot} = 5.58$. This gives a system 
luminosity $\log L/L_{\odot} = 6.13$.
 
Assuming 48~kpc for the distance of the LMC, the dust luminosity is then only 0.004 
of the stellar luminosity. Consequently, the emission is optically thin and the grain 
temperature is determined by its thermal equilibrium in the stellar radiation field. 
For grains having optical properties of amorphous carbon grains, the equilibrium 
temperature falls off with distance, $r$, from the 
stars as $T_{\rmn {g}} \propto r^{-0.38}$ (e.g. Williams et al. 2009).
In the optically thin case, the flux $F_{\lambda }$ observed at our distance 
$d$ from the system is simply proportional to the products of the grain opacity 
and Planck functions, $\kappa _{\lambda }B(\lambda ,T_{\rmn{g}}(r))$, weighted 
according to the radial mass distribution $m_{\rmn{g}}(r)$. The last is 
determined by the past history of dust formation: in the simplest case of 
constant dust formation, it would take the form $m_{\rmn{g}}(r) \propto r^{-2}$, 
but in the case of our SEDs on the rise to maximum, the gradient would be steeper, 
with relatively more hot dust nearer the stars. We therefore defer modelling the 
SEDs until we have a more complete picture of the dust formation history from 
near-IR data, together with sets of mid-IR and near-IR data more certainly at the 
same phase.

To sample the fading branch of the light curve, we formed a SED from the {\em AKARI} 
$N3$, $S7$ and $S11$ data, observed near phase 0.2 on our ephemeris. Colour corrections 
for deriving monochromatic fluxes from the in-band $S7$ and $S11$ magnitudes can be 
significant owing to the widths of the filters and we used those for a 800-K black 
body given by Lorente et al. (2008, table 4.8.11).
There are no near-IR data at this photometric phase, so we interpolated $J$ and $Ks$ 
from the ISPI and IRSF observations and show the resulting SED in Fig.\,\ref{FSEDs}. 
The black body fitted to the difference between this and the wind SED has 
$T_{\rmn bb} = 834\pm25$~K with r.m.s. residual 0.05 mag. A fit without the $S11$ 
data gives the same temperature, so we are confident in assigning the $S11$ flux 
to the WR star. Realistic modelling of this SED would require a radial 
dust density distribution to reflect the varying rate of dust formation in the 
preceeding months but it is evident from the black-body fits that the characteristic 
temperature of the dust is now significantly cooler than that of the phase 0.9 SED, 
as expected from observations on the fading branch of the light curve.

\section{Discussion}  
\label{SDiscuss}

Although the IR light curves need strengthening with further observations, they 
provide enough evidence for us to consider HD 36402 as a variable dust-making 
CWB like WR\,137. The orbit of WR\,137 (Lef\`evre et al. 2005) has periastron 
passage $\simeq 1.3\pm0.5$~yr before near-IR maximum (Williams et al. 2001), 
which is when the condensation of new grains ceased, or at least fell below the 
rate needed to replenish the dust being carried away in the stellar wind. This 
puts periastron in WR\,137 close to the middle of its 3-yr dust formation episode. 
The dust formation by HD 36402, with condensation beginning some time before IR 
phase 0.7 (cf. Fig.\,\ref{Fphased}) and ending some time near phase 0.0, looks 
very similar and suggests dust formation formed in a CWB having an elliptical orbit 
with periastron passage near our IR phase 0.85.
The candidate CWB comprises the O8I supergaint in 
an `outer' orbit with the `inner', 3.03-d WC4+O? binary, whose orbit was determined 
by Moffat et al. (1990). They found that the absorption lines attributed to the O8I 
supergaint did not share the motion of the WC4 star, suggesting that it was a third, 
more distant member of the system. We re-examined the absorption-line RVs measured 
by Moffat et al., noting that they were observed in three widely spaced runs each 
spanning 3--8 d. We determined the average RV from each run to be 283$\pm$5, 
299$\pm$8 and 241$\pm$7 km~s$^{-1}$ at IR phases 0.17, 0.41 and 0.83 respectively. 
To test whether these represented orbital motion of the O8I star about the `inner' 
WC4+O? binary, we examined the corresponding emission-line RVs measured by 
Moffat et al., adopting their orbital elements apart from the $\gamma$ velocities as 
givens, and solving for the mean $\gamma$ velocity for each run to look for motion 
of the inner binary as a single object. The mean $\gamma$ velocities for the 
three runs also showed significant dispersion, but not in anti-phase with the 
absorption-line velocities as one would expect from orbital motion, so orbital 
motion of the O8I star has yet to be determined. Further spectroscopic observations 
on a time-scale of years are needed to search for search for motion and derive an 
orbit, but here we will continue on the basis that one exists.

To estimate the properties of the `outer' (4.7-yr) CWB, we use the velocities from 
the {\em IUE} study by Fitzpatrick et al. as terminal velocities ($v_{\infty}$) 
and adopt mass-loss rates for the WC4 and O8I components of $2.5 \times 10^{-5}$ 
and $5.3 \times 10^{-6}$ M$_{\odot}$~y$^{-1}$ from those of WC4 and O8I stars 
given by Crowther et al. (2002) and Markova et al. (2004) respectively. 
These give a wind-momentum ratio 
$\eta = (\dot{M} v_{\infty})_{\rm O8} / (\dot{M} v_{\infty})_{\rm WC} \simeq 0.1$. 
Consequently, the distance between the stagnation point, where the dynamical 
pressures of the winds balance, and the O8I component will be about one-quarter 
($\surd\eta / (1+\surd\eta)$) that of the separation of the O8I and WC4 stars. 
This places the WCR close to the O8 supergaint and far from the `inner' binary, 
probably accounting for the lack of modulation of the dust emission in the {\em WISE} 
data (Section \ref{SSurveys} above) with the 3.03-d period of the latter. 

\begin{table}
\caption{X-ray observations of HD 36402 with {\em ROSAT} PSPC (rp), {\em ROSAT} 
HRI (rh) and {\em XMM-Newton} EPIC. The luminosities $L_X$ are `absorbed luminosities', 
uncorrected for extinction; those of the {\em ROSAT} observations were derived adopting 
$kT = 1.9$~keV derived from the {\em XMM-Newton} spectral fit.}
\begin{center}
\begin{tabular}{llrrcc}
\hline
Obs ID        &    Date &  Exp  &   Count rate  &  $L_X$  &    IR \\
              &         &  ks   &   ks$^{-1}$   & erg s$^{-1}$ & $\phi$ \\
\hline
rp000054a00   & 1991.33 & 4.33  &  3.15$\pm$1.03 & 1.5e34 & 0.81  \\
rp400154n00   & 1992.19 & 6.35  & -0.17$\pm$0.57 &        & 0.99  \\
rp500093n00   & 1992.19 & 8.48  & -0.21$\pm$0.49 &        & 0.99  \\
rp500054a01   & 1992.28 & 3.32  &  5.81$\pm$1.67 & 2.8e34 & 0.01  \\
rh400355n00   & 1994.20 & 4.51  &  3.51$\pm$1.31 & 5.0e34 & 0.42  \\
rh400353n00   & 1994.21 & 3.48  &  3.98$\pm$1.49 & 5.7e34 & 0.42  \\
rh400353a01   & 1994.37 & 4.41  &  2.80$\pm$1.21 & 4.0e34 & 0.46  \\

0071940101 & 2001.83 & 27.57 & 15.9$\pm$0.9   & 3.0e34 & 0.04  \\
\hline
\end{tabular}
\end{center}
\label{TXray}
\end{table}

Bomans et al. (2002) drew attention to the high X-ray luminosity 
($L_X = 9 \times 10^{35}$ erg~s$^{-1}$ of HD 36402 measured with {\em XMM-Newton}, 
but noted also the softness of the spectral index ($\Gamma = 2.6$), 
which argued against its being caused soley by collision of the stellar winds.
X-ray emission by HD~36402 was also observed using the {\em ROSAT} Position 
Sensitive Proportional Counter (PSPC) and High Resolution Imager (HRI) with 
0.5--7.0~keV luminosities of $5.4\pm1.4 \times 10^{33}$ erg~s$^{-1}$ and 
$3.2\pm0.9 \times 10^{34}$ erg~s$^{-1}$ respectively (Guerrero \& Chu 2008). 
The individual {\em ROSAT} observations were recovered and are given in 
Table \ref{TXray}. To facilitate comparison, luminosities were derived from 
the {\em ROSAT} count rates using web~{\sc pimms} 
and adopting the absorption column $N_H = 3.4 \times 10^{20}$~cm$^{-2}$ and 
$kT = 1.9$~keV derived from the {\em XMM-Newton} spectral fit, i.e. folding
all the variability into the `absorbed luminosities' $L_X$.

It is apparent that the X-ray flux observed from HD 36402 is variable. 
Most striking are the non-detections in the 1992 {\em ROSAT} PSCP observations 
in 1992.19 although the integration times were longer than those before and 
afterwards, in which HD 36402 was detected. The 1992.19 observations occurred 
close to zero IR phase, and perhaps $\simeq 0.15P$ after periastron passage in 
the 4.7-yr binary believed to be responsible for the variable dust formation. 
The non-detection of X-ray emission from HD 36402 followed by its quick recovery 
invites comparison with the narrow minimum in the X-ray light curves of WR\,140, 
which occurs just after periastron passage (e.g. Pollock et al. 2005). 

The X-ray flux observed from a CWB depends on both the intrinscic emission measure 
and the photoionization extinction through the stellar winds, especially the heavier 
wind of the WC star. The source is expected to lie close to the stagnation point 
between the WC4+O? and O8I winds and, as seen above, much closer to the O8I star. 
The circumstellar extinction varies as the orbit progresses and the source moves 
through the WC4 wind according to the orbital elements -- expressions are given 
by Williams et al. (1990) -- and is usually greatest near periastron, when the source 
is most deeply buried in the WR stellar wind, or conjunction, when the source passes 
behind the WR star. The non-detection of X-ray emission from HD 36402 at phase 0.99 
and relative weakness at phase 0.81 are consistent with this picture but modelling  
the X-ray variations requires knowlege of the `outer' orbit. 
The intrinsic emission measure is also expected to vary round the orbit, as $1/D$ 
where $D$ is the separation of the stars, for an adiabatic WCR 
(Stevens, Blondin \& Pollock 1992). This would make the intrinsic emission measure 
somewhat greater at periastron, but without more spectroscopy or at least 
measures of the hardness ratio around the orbit, we cannot tell at this stage.

The relatively wide separation of the outer binary implied by its period should 
allow non-thermal radio emission from the WCR to emerge (cf. Dougherty \& Williams 2000), 
at least away from periastron, but the distance of the LMC would make this a 
challenging observation with current instrumentation -- e.g. the maximum 6-cm flux 
density of WR\,140 shifted to the distance of the LMC would be $\sim$ 30~$\umu$Jy.

The demonstration that one of the WR systems in the LMC is an episodic dust-maker 
like those in the Galaxy suggests that there are more to be found. 
Other candidate dust-variables include HD~32125 (BAT99--9) and HD~38448 (BAT99--125).
In the case of HD~32125, most of dispersion in $Ks$ comes from the DENIS observation, 
which has a relatively large uncertainty ($\sigma$ 0.22 mag), so it would be premature 
to consider it a episodic dust-maker and both stars need re-examination for variability 
when photometry from the VMC survey becomes available. If episodic dust formation is 
possible in the low metallicity environment of the LMC, what about persistent dust 
formation, like that observed from some WC8--9 stars observed in the inner regions of 
the Galaxy? One candidate amongst the LMC WR systems is BAT99--121, which has 
anomalously red $J$--$Ks$ colours in both the 2MASS and IRSF MC surveys. This star is 
in a difficult field and will be discussed elsewhere. In any event, HD~36402 has posed 
the question: why has dust formation not been observed from Galactic WR systems outside 
the solar circle, where the overall metallicity is lower?
A second question is what is the lowest metallicity environment in which WR systems can 
make and contribute dust to the ISM?

\section*{Acknowledgements}

This work is based in part on observations made with the Spitzer Space Telescope, 
which is operated by the Jet Propulsion Laboratory, California Institute of 
Technology under a contract with National Aeronautics and Space Administration 
(NASA). This work also makes use of data products from the AKARI mission, 
a JAXA project with the participation of ESA,  and the Wide-field Infrared 
Survey Explorer, which is a joint project of the University of California, 
Los Angeles, and the Jet Propulsion Laboratory/California Institute of 
Technology, funded by the National Aeronautics and Space Administration. 
This research has made use of the NASA/IPAC Infrared Science Archive, 
which is operated by the Jet Propulsion Laboratory, California Institute 
of Technology, under contract with the NASA; the Vizier database, operated by 
the CDS, Strasbourg, and the DARTS archive developed and maintained by C-SODA at 
ISAS/JAXA.
PMW is grateful to the Institute for Astronomy and UK Astronomy Technology Centre 
for continued hospitality and access to the facilities of the Royal Observatory Edinburgh.

\end{document}